# Tuning the work function of graphene toward application as anode and cathode

Samira Naghdi[1,*], Gonzalo Sanchez-Arriaga[1], Kyong Yop Rhee[2,**]

[1]Bioengineering and Aerospace Engineering Department, Universidad Carlos III de Madrid, 28911, Leganes, Spain

[2]Department of Mechanical Engineering, College of Engineering, Kyung Hee University, 446-701, Yongin, Korea

[*]Corresponding author. Tel: +34 631 817 573, snaghdi@ing.uc3m.es (S. Naghdi)

[**]Co-corresponding author. Tel: +82 31 201 2565, rheeky@khu.ac.kr (K. Rhee)

## Abstract

The rapid technological progress in the 21st century demands new multi-functional materials applicable to a wide variety of industries. Two-dimensional (2D) materials are predicted to have a revolutionary impact on the cost, size, weight, and functions of future electronic and optoelectronic devices. Graphene, which shows potential as an alternative to conventional conductive transparent metal oxides, may play a central role. Since its work function (WF) is tunable, graphene exhibits the interesting ability to serve two different roles in electronic and optoelectronic devices, both as an anode and a cathode. After introducing some basic concepts, this work reviews the most important advances in controlling the tuned WF of graphene, highlighting special features of graphene electronic band structure and recognizing different methods for measuring WF. The impact of thickness, type of contact, chemical doping, UV and plasma treatments, defects, and functional groups of graphene oxide are considered and related with the applications of the modulated material. The results of the review, organized in lookup tables, have been used to identify the advantages and main challenges of the tuning methods.

## 1. Introduction

Synthesis of graphene for various applications has become a very popular trend in the last decade since graphene exhibits great potential as an alternative to conventional conductive transparent metal oxides such as indium tin oxide (ITO). Graphene may be produced from an abundant and cost-effective precursor (such as graphite) [1-4]. Graphene presents essential mechanical properties such as flexibility, stiffness, and robustness, along with high thermal and electrical conductivity for electronic devices and good transparency as demanded by optoelectronic applications [5-8]. These extraordinary characteristics are due to the 2D structure composed of $sp^2$-bonded carbon atoms, in which a small structural defect would drastically change the properties of graphene.

One of the reasons for the popularity of graphene in the electronic and optoelectronic industries is the adaptability for new applications. Graphene presented the ability to be used as a hole injection material and as an electron injection electrode. These properties are a consequence of the special electronic band structure of graphene. Unlike insulators, whose energy bands are either full or empty and are separated by an energy gap, and conductors, which have a partially filled band, the energy-momentum diagram of graphene has two circular cones connected at one point (Dirac point, Figure 1a). Therefore, there is no gap nor a partially filled band. This work is focused on methods for controlling the difference between the Fermi level ($E_F$) and the energy of an electron at rest in a vacuum immediately outside the solid surface ($E_v$). This difference ($E_v$- $E_F$) is called work function (WF) (Figure 1), which is also defined as the minimum energy needed to remove an electron from a metal surface. The WF is not a fundamental characteristic of a bulk material but a property of its surface.

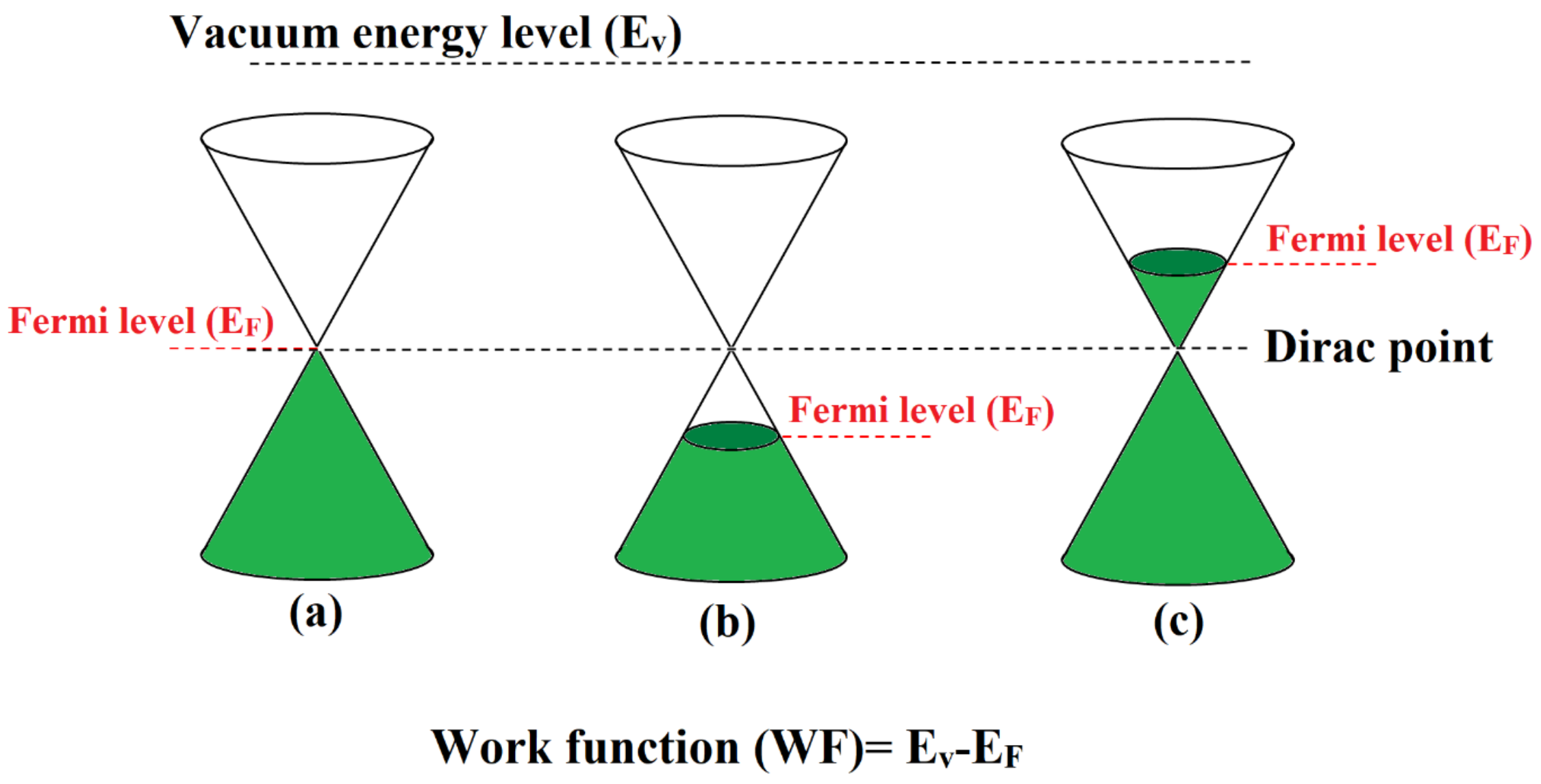

**Figure 1.** Schematic diagram of energy band of (a) pristine graphene, (b) increasing the WF of graphene, (b) decreasing the WF of graphene.

Based on this definition, shifting the Fermi level of graphene via an applied electric field, surface engineering, doping, etc. would manipulate the WF (Figure 1b,c) [9, 10]. The WF of graphene is about 4.6 eV (in the range of graphite), for which its application as a cathode must be decreased, while use as an anode requires a high WF (to increase the hole injection efficiency) [11]. Therefore, using graphene in two different electronic parts demands some modifications in its electronic behavior [6, 12-16].

For application as cathodes in electroluminescent devices, thermionic energy converters (TECs), hollow-cathodes, photocathodes, photon-enhanced thermionic emission (PETE) devices [17], and low work-function space tethers [18, 19], modulation methods are needed to decrease the WF. A small modulation of the WF typically produces a strong variation in the performance of these devices. For instance, the density current $J$ emitted by a sample of work function $W$ at temperature $T$ due to the thermionic effect follows the Richardson-Dushmann law [20]:

$$J = \lambda A_0 T^2 \exp(-W/k_B T) \quad (1)$$

where λ is a material-specific correction factor, $A_0 = 4\pi m_e k_B^2 e/h_p^3$, $k_B$ is the Boltzmann constant, $e$ is the elementary charge, $m_e$ is the electron mass, and $h_p$ is Planck's constant. Therefore, the emitted current varies exponentially with $W$, and modulations of the WF in tenths of an electron volt (eV) can impact performance. Such a sensitivity is also found when a device operates through the photoelectric effect.

Furthermore, in the application of graphene in organic field-effective transistors (OFETs), organic light emitting diodes (OLEDs), and organic photovoltaic cells (OPVs) as both anode and cathode, not only are the transparency and conductivity of electrodes important, but the contact between the semiconductor and electrodes needs special attention. The stability and performance of such optoelectronic devices are dependent on both the properties of the active materials (semiconductors) and on the interfacial properties (contact resistance) of the semiconductor/electrodes. The electrodes in these devices either inject or extract an electron from the semiconductor. Any mismatch of the WFs between organic semiconductors and electrodes results in high contact resistance, which decreases the charge injection and extraction efficiency [21-23]. Therefore, clever design of graphene with both high electron injection (low WF) and high hole injection (high WF) abilities is in demand [24-29].

Since the application of graphene as an electrode in future electronic and optoelectronic industries would make a dramatic change in the industry (weight, size, price, mechanical properties, etc.), research effort has been performed in the last decade to develop modulation methods for tuning the WF of graphene. The goal of this work is to present the main results in a structured and coherent manner to highlight state-of-the art tuning methods, limitations, and challenges to be addressed in the near future. Before presenting tuning methods, Sec. 2 reviews the most important techniques for measuring WF. This preliminary step is necessary because each diagnostic technique exhibits its own peculiarities, which should be kept in mind to interpret the results of the tuning methods. The latter are thoroughly reviewed in Sec. 3 that considers the impact of thickness, type of contact, doping, UV and plasma treatments, defects, and functional groups of graphene oxide on WF. Lookup tables are included to ease the search of information as well as provide a summary of the acronyms. Finally, the main conclusions regarding the different tuning methods and some ideas for future work are presented in Sec. 4.

## 2. Measuring the work function of graphene

A good understanding of the tuning methods requires a brief discussion regarding the most important diagnostic techniques for measuring the WF. It is not a minor subject, as shown by the different WF values of materials in the literature. An example is the C12A7:$e^-$ [30], one of the most promising electrodes due to its extraordinary stability and low WF [31]. Using data from field-emission and thermionic-emission characteristics, photoelectron yield spectroscopy (PYS), and UV photoelectron spectroscopy (UPS), the values 0.6 eV [32], 2.1 eV [33], and 2.4 eV [34] were determined. A space group working on hollow cathodes estimated a value of 0.76 eV from experimental results [35].

Although the values found in the literature for graphene are much more uniform, measuring the WF without a standard device and accurate method could result in different WF values for the same material. The two dominant diagnostic methods for graphene are Kelvin probe force microscopy (KPFM) and UPS, a few others such as thermionic emission and field emission are reviewed because they are connected with interesting applications. This brief introduction to diagnostic methods assists in identification and selection of the appropriate techniques, highlighting the advantages and drawbacks. Contamination on the surface and ambient humidity affect the measurements, and high vacuum conditions are recommended.

### 2.1. Kelvin probe force microscopy (KPFM)

One of the most accurate and widely used methods for measuring the WF of graphene is Kelvin probe microscopy (SKPM or KPFM), which is an experimental technique based on an atomic force microscope (AFM). As shown in Figure 2, KPFM, also known as surface potential microscopy, has a cantilever with a metallic tip, an applied electric modulation ($V_{ac}$), and a controller that adjusts the bias $V_{dc}$ to compensate for potential differences between the tip and the sample. If the WF of the tip ($W_{tip}$) is known, the WF of the sample ($W_s$) can be calculated from

$$W_s = W_{tip} - eV_{dc} \qquad (2)$$

Interestingly, the WF difference between two points of the sample does not depend on $W_{tip}$ and is given by $\Delta(W_s) = - e\Delta(V_{dc})$ [36, 37]. Therefore, a map of the WF and information regarding the electronic state and composition of the sample surface and its homogeneity are obtained. Moreover, while measuring the WF, the topographic map with correct step heights of the surface potential of the sample provides thickness data for the graphene layers. However, since electrostatic forces between the sample and cantilever tip determine the WF, this method is very sensitive to the measurement environment (such as humidity). The stability of its measurements of $W_s$ also depends strongly on the stability of the WF of the cantilever tips [38].

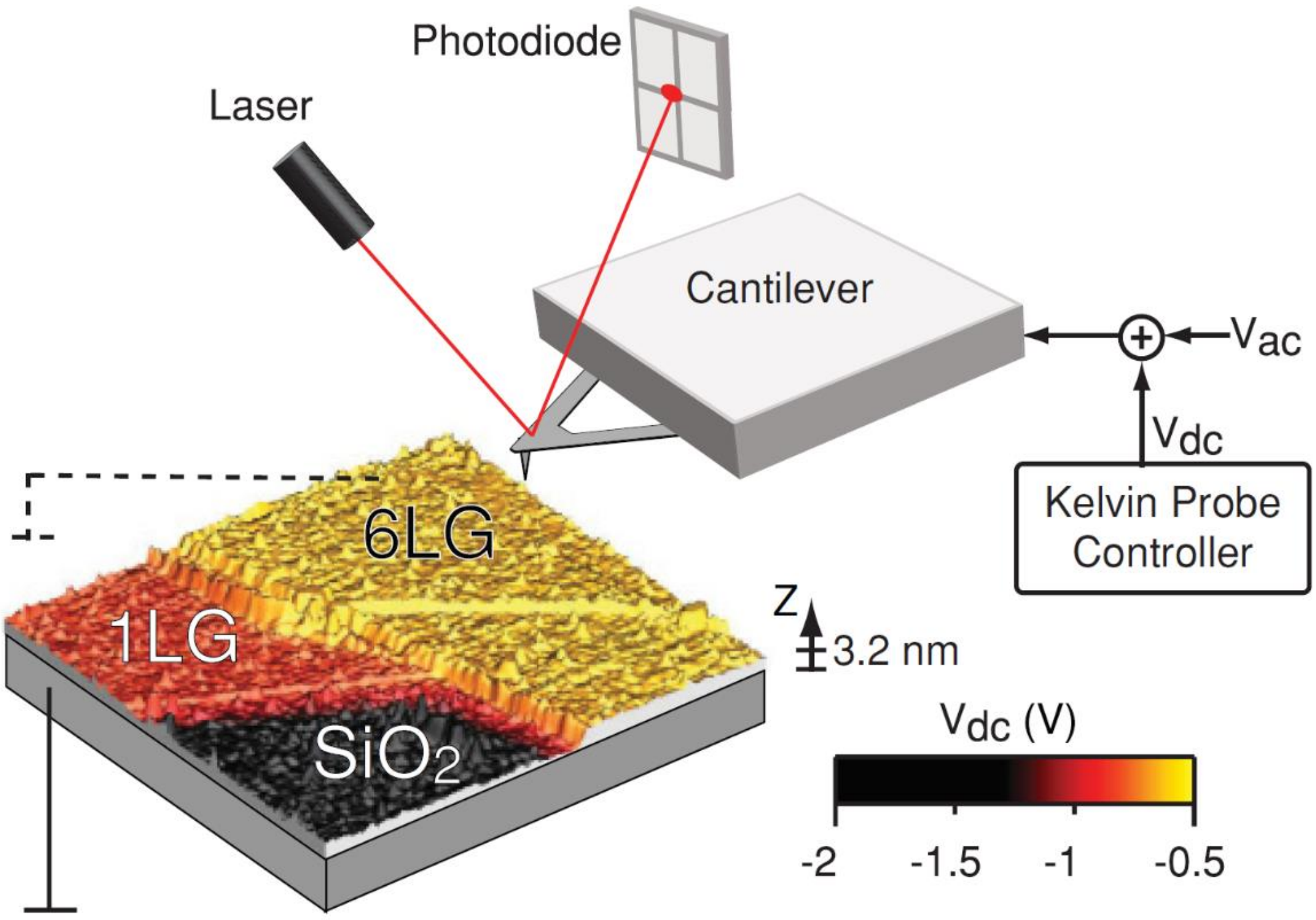

**Figure 2.** Schematic illustration of KPFM setup for measuring the WF of one-layer (SLG) and 6-layer graphene (FLG) on $SiO_2$ substrate ($V_{dc}$ is the contact potential difference). Reprinted with permission from Ref. [39].

### *2.2. Ultra-violet photoelectron spectroscopy (UPS)*

UPS determines the WF of a sample illuminated with ultraviolet photons by measuring the kinetic energy spectra of the emitted photoelectrons. This method involves a spectrometer equipped with a hemispherical energy analyzer and an excitation source such as a He discharge lamp. The WF is determined by subtracting the width of the photoelectron spectrum from the photon energy of the excitation light (hν). The former is calculated from the spectrum as the difference between the Fermi edge ($E_F$) and the inelastic high binding energy cutoff ($E_{cutoff}$). To calibrate the position of the Fermi level, the Fermi edge of gold is normally measured. The latter is accurately known, and depending on the operation conditions, UV sources of He I (hν =21.22 eV) or He II (hν =44.8 eV) can be used. The WF of the sample $W_s$ is then computed as

$$W_s = h\nu - \left(E_F - E_{cutoff}\right) \quad (3)$$

UPS is a surface-sensitive method, and existence of contamination on the surface would drastically affect the measured WF [40]. The measurement is normally performed in ultrahigh vacuum conditions.

### *2.3. Thermionic emission*

Unlike UPS, where the photons provide the energy for electrons to overcome the WF of the material, thermionic emission is a thermally-excited charge emission process. The sample is heated, providing thermal energy to the charge carriers (electrons or ions) and helping them to overcome the WF. This phenomenon follows the Richardson-Dushmann equation [Eq. (1)], which relates the emitted current density of the surface to the WF of the sample. Therefore, if the emitted electrons are collected by an anode, an electric current is produced, and the work function can be determined from Eq. (1). However, this method is subject to different sources of errors. For instance, field effects may affect the method if a potential difference above a certain threshold is set between the sample (cathode) and the anode. Also, at high temperature (strong emission), an electric field that would attract some electrons (the less energetic) back to the probe is developed. Due to this space-charge effect, a current density below that calculated by Eq. (1) would be measured. The measurement at high temperature requires models to distinguish the different operational regimes [41]. On the other hand, measuring the WF at high temperatures is advantageous because the adsorbent would be removed from the surface of the sample, so the presence of contamination on the surface would not affect its WF. Graphene, as a material that can withstand high temperature, is heated to incandescence until an obvious thermionic emission current can be detected [42].

### *2.4. Field electron emission characteristics*

In field electron emission, also known as cold emission, electrons are extracted from the surface of the sample by an electrostatic field $E$. The current density is approximately described by the Fowler-Nordheim model

$$J = A\frac{(\beta V)^2}{\alpha W}exp\left(-\frac{BW^{3/2}}{(\beta V)}\right) \quad (4)$$

with $A = 1.514 \times 10^{-6} AeVV^{-2}$, $B = 6.831 \times 10^{-7} eV^{-3/2} Vcm^{-1}$, and the field $E$ and the bias $V$ related by $E = \beta V$, with $\beta$ as a local field conversion factor. After measuring the current-voltage characteristic (I-V), a Fowler-Nordheim plot with $\ln(I/V^2)$ versus $1/V$ can be fitted to a straight line. The slope and cut with the $\ln(I/V^2)$-axis provide information about the field conversion factor, the work function, and the effective emission area. However, electron emission can be affected by factors beyond the simple model used to derive Eq. (4) that, in principle, is only appropriate for bulk materials. For that reason, the WF obtained with Eq. (4) is not intrinsic and can depend on the geometry of the emitter, as well as its electronic structure. The value of the WF is also sensitive to possible contamination of the emitter surface.

### *2.5. Photoemission electron microscopy (PEEM) or photoelectron microscopy (PEM) images using secondary electrons*

PEEM is a type of emission microscopy that generates image contrast by utilizing the local variations in electron emission. To excite the electron, sources of energy such as X-ray radiation, UV light, or synchrotron radiation are used. Since the electron emission is from a shallow layer, the surface of the sample is very sensitive. This method is used for flat and smooth surface samples to eliminate topographic effects and to enhance lateral resolution. In this method, variation of the electron emission from different parts of the surface (with a different WF) results in image contrast. From the PEEM images composed of secondary electrons, areas with different numbers of graphene layers could be recognized. By increasing the number of graphene layers, secondary electron emission shifts to higher energies, which can be observed in the secondary electron emission spectra [43, 44].

### *2.6. Other methods*

Internal photoemission (IPE) spectroscopy [45], low-energy electron microscopy (LEEM) [44, 46], scanning tunneling microscopy (STM) [47], scanning photocurrent microscopy (SPCM) [9], and capacitance-voltage (C-V) measurements using a metal-graphene semiconductor capacitor structure [48, 49] are other methods used worldwide for measuring the WF of graphene.

## 3. Modulation of the work function of graphene

Although the WF of a freestanding graphene layer is about 4.6 eV, there are factors that could affect the WF, and several reported values are different. Certain factors are related to the structure of graphene itself (thickness, functional groups, defects, disorder), and other factors are related to the application of graphene (substrate and buffer layer, chemical doping, UV and plasma radiation). By analyzing these factors, tuning the WF of graphene would present new opportunities for investing in this versatile material and taking advantage of its extraordinary properties.

### *3.1. Thickness dependence of the graphene work function*

The thickness of graphene, i.e., the number of layers, is one of the main factors that affect the WF. It also modifies other key characteristics such as heat and electrical conductivity and transparency. Based on the number of layers, graphene can be divided into three groups; single-layer graphene (SLG, 1-layer graphene), bi-layer graphene (BLG, 2-layer graphene), and few-layer graphene (FLG, between 2-10 layers of graphene). Therefore, choosing the most accurate method for producing graphene and good control over the graphene layer deposition is very important. In this regard, important research effort has been performed to study the relationship between graphene thickness and WF. The WF of graphene with more than one layer is known to be dependent on the number of layers due to the charge transfer at the substrate interface and the charge distribution within the different layers of graphene [50]. The results of PEEM studies revealed that, by increasing the number of graphene layers, the C1s core level shifted to lower binding energies as a result of the thickness dependence of the Dirac point energy (Figure 3a) [44].

Filleter et al. found that the WF of SLG films (grown epitaxially on 6H-SiC(000)) measured by KPFM was 135.9 meV lower than the WF of a BLG [51]. Later studies reported the same results [10, 37, 38, 43, 44, 52] and indicated that the SLG could be chemically less stable than the BLG and FLG [53]. Panchal et al. studied the surface potential and the WF of the SLG and BLG domains via a combination of electrical functional microscopy and spectroscopy techniques [54]. They reported values of 4.55 eV and 4.44 eV for SLG and BLG, respectively, in ambient conditions. Based on their results, since WF is a surface characteristic rather than a bulk property, environmental conditions could drastically affect the graphene WF. Analysis of the thickness dependence of WF of FLG revealed that saturation occurs at five layers of graphene due to interlayer screening [49]. The WF of FLG from three to ten layers is independent of the number of layers and is about 4.43 eV (Figure 3b) [55].

Research on the thickness dependency of the WF of reduced graphene oxide (rGO) had the same results [16]. The possible reason for increasing the WF of rGO with the number of layers was attributed to the oxygen concentration between the rGO layers. Misra et al. showed that the oxygen concentration in rGO layers could be controlled by reducing the GO sheets and changing the rGO thickness [16].

The layer-by-layer transfer of the chemical vapor deposition (CVD) grown graphene on a Si substrate and photovoltaic effects of a heterojunction structure embedded with FLG were investigated [56]. They demonstrated dependence of the open-circuit voltage ($V_{oc}$) on the number of graphene layers. The results of their investigation proved the dependence of the WF of FLG on the number of layers, which could modulate the photovoltaic behavior of FLGs/Si interfaces [56]. Misra et al. investigated the thickness dependency of the graphene WF using different graphene thicknesses as a metal gate electrode in a metal oxide semiconductor structure by inserting them between the dielectric ($SiO_2$) and contact metal (TiN) [57]. The results showed that the effective WF of the gate electrode could be tuned up to 0.5 eV by varying the number of graphene layers. Functional improvement of the metal oxide semiconductor was ascribed to the impermeability of graphene for TiN, which resulted in reduced metallic contamination in the dielectric [57].

Doping graphene layers with different thickness was investigated both numerically (DFT) and experimentally (KPFM) [37]. It was shown that doping the SLG and BLG shifted the Fermi energy with respect to the Dirac point and led to a variation of WF, while interlayer screening was the reason for the variation in WF for FLG. Moreover, doping the FLG with more than four layers negated the dependence on the number of layers.

Contacts between the graphene layer and different materials, a topic that will be addressed in the next section, affects its WF. The interesting finding here is that the effect of metal contact on the WF of graphene varied for graphene layers with different thicknesses. Song et al. reported that tuning the WF of graphene in contact with metal from 4.3 eV to 5.1 eV could be achieved by changing its thickness [48]. They showed that the WF of SLG was easily affected by the WF of the metal contact, while that of FLG was less sensitive (Figure 3c). Therefore, not only is the WF of graphene dependent on the number of layers, but also the doping behaviors of metal contacts and chemical dopants are affected by the number of layers.

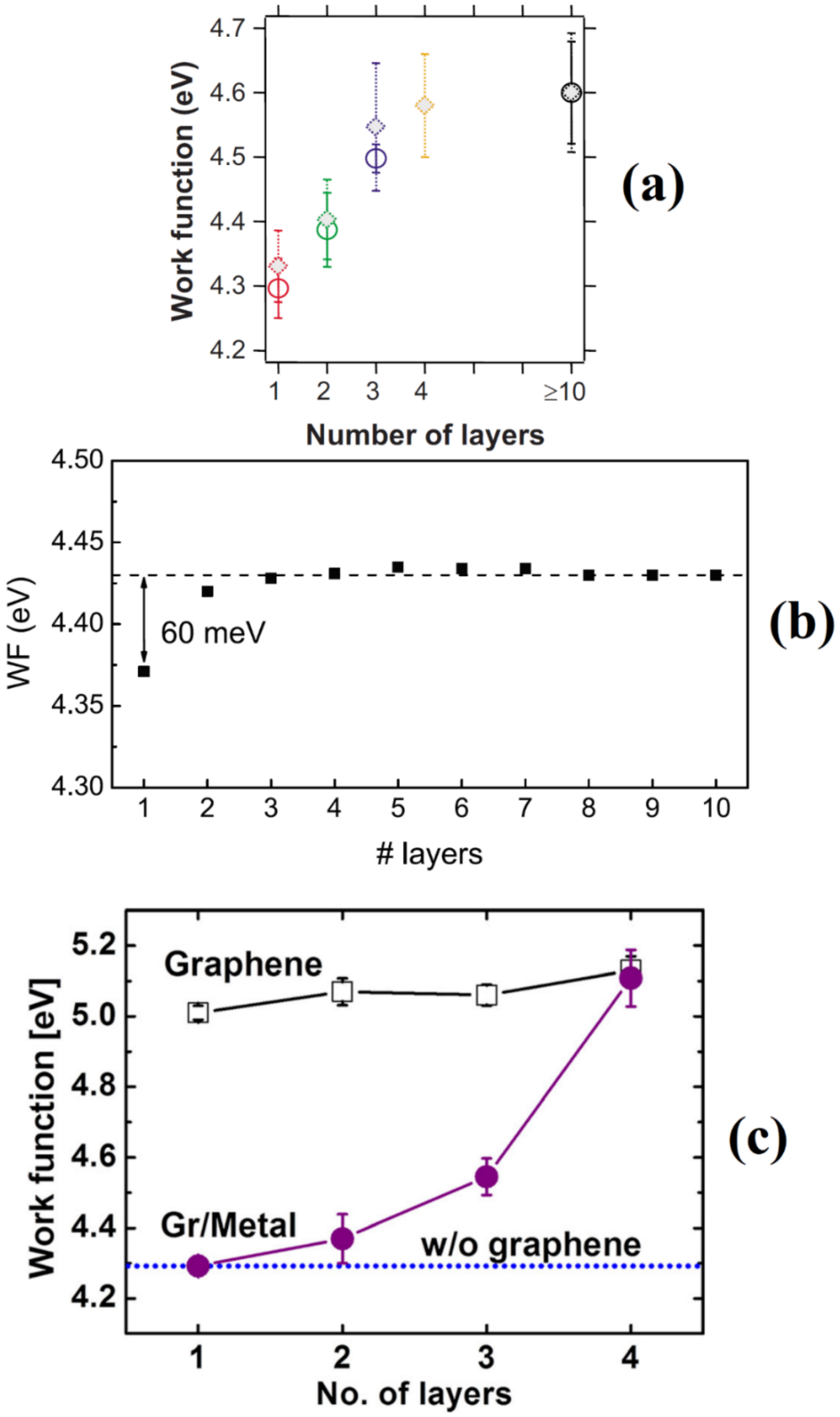

**Figure 3.** Panels (a), (b), and (c) show WF of graphene versus the number of layers. Reprinted with permission from Ref. [58], [59], [60].

### *3.2. Effect of the type of graphene contact on its work function (contacted metal or insulator)*

The WF of a free-standing 2D sheet of graphene has been calculated theoretically [61]. However, physical contact of the graphene with the environment exists in real applications, and impurities or an insulator substrate should be considered. A graphene layer could exist on a substrate (graphene coating), be coated with a buffer layer, or contain many types of contact that affect the WF [14, 47, 62]. In theory, the Fermi level of pristine graphene is considered to coincide with the Dirac point (Figure 1a), while in real scenarios, graphene contact with a metal (or insulator) alters its electrical structure. Therefore, a full understanding of the physics of the metal (insulator)/graphene interfaces is very important.

Upon contact of graphene with a metal (insulator), the WF difference results in transfer of the charge to create an interface dipole moment, which controls the WF of a graphene layer [9, 50, 52]. The dipole size and the sign and magnitude of doping in a graphene layer depend on the type, thickness, and WF of the contact. Based on the strength of the adsorption of the graphene layer to its contact layer, the electronic structure of the graphene would be affected. The strong chemisorption would change the electronic structure of graphene, while weak binding physisorption transfers the electron from (to) the contact and moves the graphene at Fermi level upward (downward) from its original state [61].

For instance, transferring an electron from graphene to a modified SiC substrate (with F4-TCNQ as the electron acceptor element) resulted in p-doped graphene [63, 64]. This epitaxially grown graphene on modified SiC showed an increase in WF of about 0.7 eV compared to graphene on unmodified SiC. The results revealed that, by increasing the F4-TCNQ thickness, the WF of graphene increased further. The interfacial charge transfers between graphene and F4-TCNQ resulted in an upward shift of the vacuum level (increasing the WF) that increased the potential application of graphene used in the nanoelectronics industries [63]. Growing graphene on a SiC substrate would also result in changing the Fermi level of graphene and the WF. Filleter et al. showed that the Fermi level of the free-standing graphene coincides with the energy of the Dirac crossing point, while epitaxially grown graphene would show a shifted Fermi level [51]. This shift was in accordance with the n-doping of the graphene layer caused by the SiC substrate. The charge transfers across the interface and the impurities in the SiC substrate were explained as the reasons for n-doping of the graphene [51].

The strength of graphene adsorption on a metal contact may be divided into two groups [65]. Chemisorption occurs when graphene is in contact with metals such as Co, Ni, and Pd, and its electronic structure is drastically affected by these metallic contacts. In contrast, graphene is weakly adsorbed (physisorption) in metallic contact with Au, Al, Cu, Pt, and Ag. In both cases (chemisorption and physisorption), graphene was doped with electrons or holes, and its Fermi level was moved from the Dirac points. Contact between graphene and Al, Ag, and Cu would produce n-doped graphene (upward Fermi level shift), while Au and Pt contacts result in a p-doped graphene layer (downward Fermi level shift) (Figure 4a,b) [61]. Similar results were reported independently [55]: the type of metal contact affects the type of doping and the direction of the Fermi level shift. Graphene is chemisorbed on Co, Ni, Pd, and Ti (strong bonding), and its WF is reduced (n-doped), in which this strong interaction considerably perturbs the electronic structure of graphene. In contrast, for graphene physisorbed on Al, Cu, Ag, Au, and Pt surfaces (weak bonding), the interaction between graphene and the metal contact occurs over a short range. Moreover, the graphene was p-doped (n-doped) for contact with metals having a WF larger (smaller) than 5.4 eV [61].

To investigate the impact of metal contact on the WF of graphene, another group investigated graphene WF modulation with different metals [49]. The results of their investigation revealed that the WF of the graphene was dependent on the WF of the metal contact (as it was reported by other groups). If the contacted metal was Cr/Au or Ni, the WF of graphene was pinned to the metal, while for a Pd or Au contact, the WF of graphene was about 4.62 eV (regardless of the WF of the contact metal). They predicted that the stronger interaction between graphene and Cr/Au or Ni was responsible for changing the graphene WF as its unique characteristic, while the interaction between graphene and Pd or Au is not strong enough to produce such a change. Furthermore, they found that increasing the WF of graphene could not reduce the contact resistance, so the WF was not the only function for determining the contact resistance [49].

Yi et al. showed that even a thin layer of metal contact (Al layer less than 0.6 nm thick) could drastically change the WF of graphene [66]. They deposited a thin layer of Al on FLG and investigated the effective WF of graphene via the UPS method. The results revealed that the charge transfer between FLG and a thin Al layer equilibrated the Fermi level, changing the effective WF of FLG from 3.77 to 4.40 eV. This modulated WF of FLG may be suitable for applications in the electronics and optoelectronics industries (Figure 4c). Jo and his group showed that engineering of the graphene WF would facilitate its application both as a cathode and anode in OSCs [15]. Coating the graphene electrode with high WF materials (such as PEDOT:PSS) enhanced the hole extraction (anode), while the WF of graphene should be decreased for use as a cathode in inverted-structure OSCs. In their research, an interfacial dipole layer of WPF-6-oxy-F was used to enhance the device performance by improving charge extraction and its application as a cathode. The interfacial polymer layer controlled the WF of graphene by its ionic or polar groups. Regarding the application as an anode, they successfully produced a decrement in the WF of graphene of 0.05±0.03 eV, 0.22±0.05 eV, and 0.33±0.03 eV using PEO, $Cs_2CO_3$, and WPF-6-oxy-F, respectively (Figure 4d,e) [15].

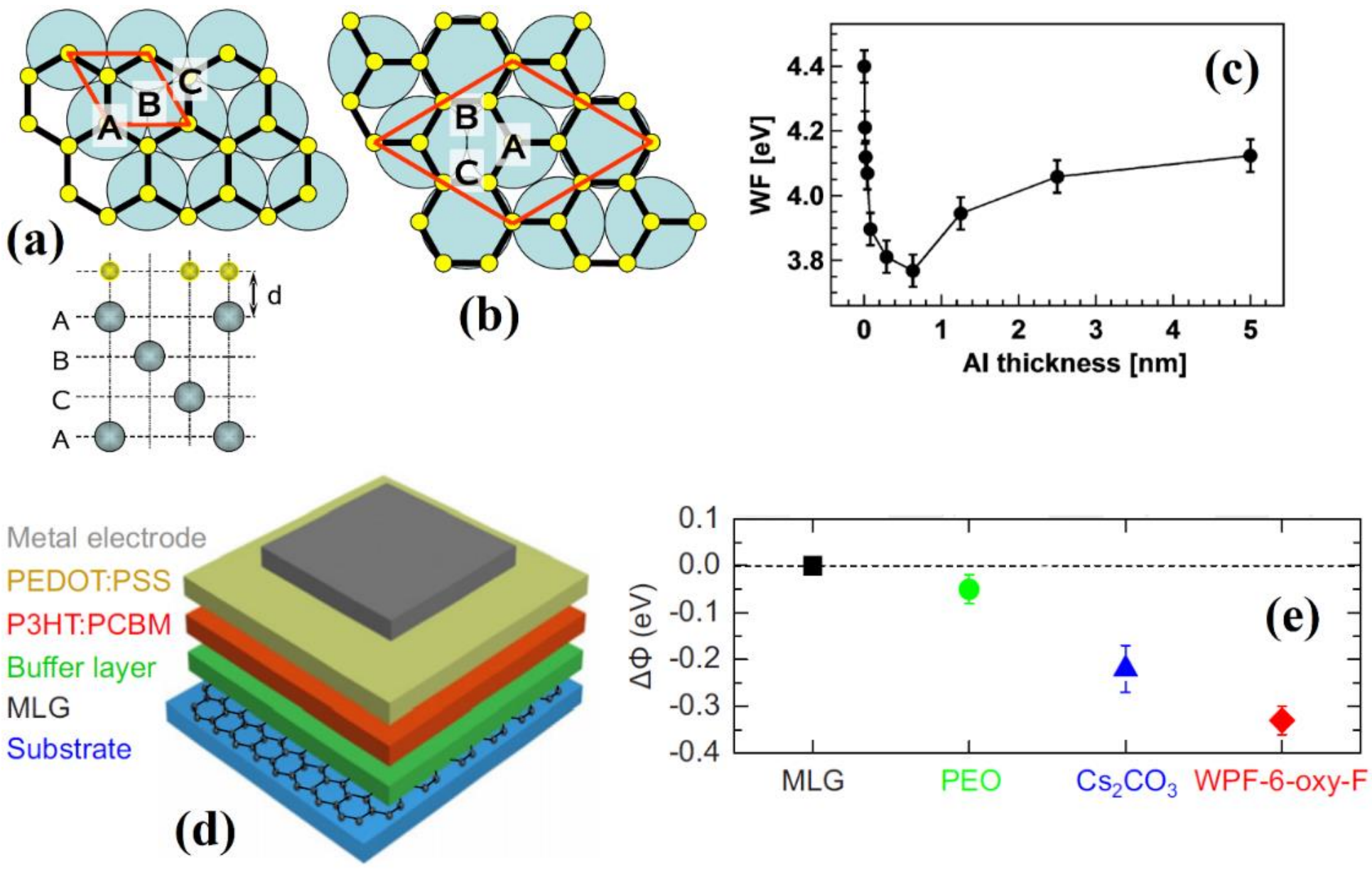

**Figure 4.** (a) The most stable structure of graphene on Cu, Ni, and Co (111) and (b) on Al, Au, Pd, and Pt (111), (c) impact of substrate thickness on WF of graphene, (d) schematic of a OSC with a FLG (WF modulate with a buffer layer) as electrode, and (e) dependence of effective WF of graphene with different buffer layer. Reprinted with permission from Ref. [67], [68], [69].

Misra et al. reported the WF of rGO under different capping metals (Pt, Ir, Al, TiN) [16]. They found that, in the case of Pt, Ir, and Al, the WF of the rGO mainly depends on the thickness of the rGO layer and not on the type of capping metal. For example, the WF of TiN affected the WF of the tuned rGO. A possible explanation relates to the type of interaction of the rGO with Pt, Ir, and Al (physisorption) and its interaction with TiN (chemisorption) [16].

In certain scenarios, it is interesting to contact the graphene with other materials without changing its WF. In this regard, Zhu et al. reported CNT film as a perfect substrate for the graphene layer for preserving its intrinsic WF [42]. A super-aligned CNT film was utilized as a porous support for graphene, and the WF of graphene was measured via a thermionic emission method. They showed that, based on the porous nature of the CNT network, the substrate influence on the WF of graphene could be minimized. Due to the high temperature sustained during the thermionic emission, the influence of adsorbents was excluded, and the WF of graphene was about 4.74 eV. The adsorbent-free measurement conditions and the weak interaction between the CNT support and graphene, due to the porous nature of the CNT substrate, suggested that the measured WF in this method was the intrinsic WF of graphene (Figure 5a,b) [42].

Another interesting factor that can impact the application of exfoliated graphene in the industry is the size and configuration of graphene flakes. Flakes affect the interaction with the substrate and produce inhomogeneity of the WF of the graphene layer. Abdellatif et al. showed that overlap of small graphene flakes, accumulation of electrical charge at the flake edges, and interaction between graphene flakes and the substrate would result in electronic noise [70]. To render the effect of WF inhomogeneity of the graphene layer, exfoliated graphene was deposited on a rough Ag substrate by a dip coating method, and the graphene layer was studied by SKPM. The results showed that engineering the substrate minimized the WF inhomogeneity of the graphene layer. The interaction between graphene and its substrate determined the electronic noise in the optoelectronic devices by controlling the charge accumulation (Figure 5c,d).

Regarding humidity, its existence in ambient conditions cannot affect the WF of a graphene layer deposited on a charge-donating substrate unless the water layer is placed between the graphene and the substrate. In that case, the water layer can change the charge doping considerably [55]. From the above information, it can be concluded that the WF of graphene is modulated when it is contacted with other conductive or insulating materials of varying WF and thickness and by the type of the interaction between the graphene layer and contacted material (physisorption or chemisorption).

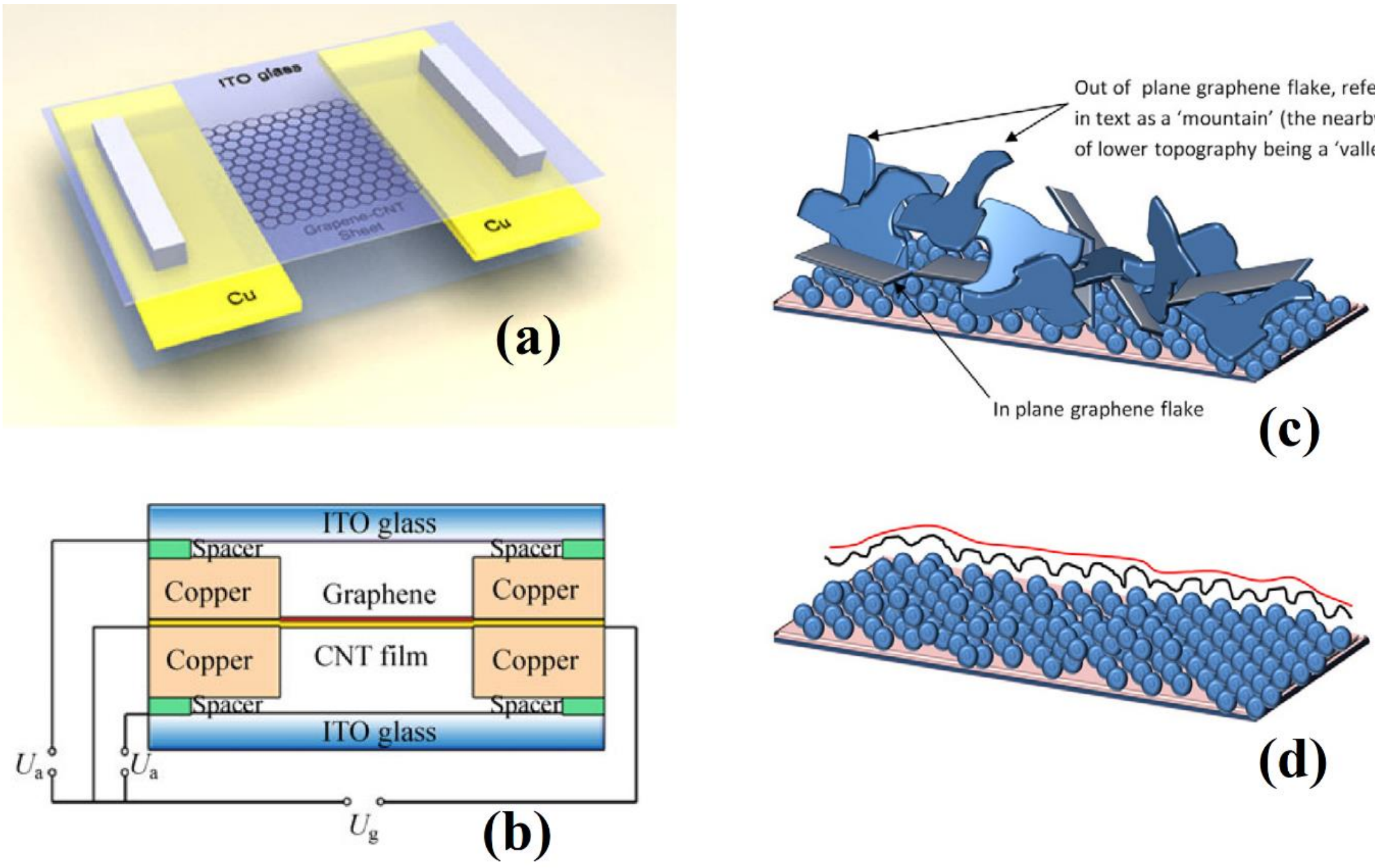

**Figure 5.** Schematic illustration of (a) thermionic emitter, and (b) its electronic circuit, (c) schematic diagram of graphene flake on a rough surface, and (d) decreasing the surface roughness by depositing a SLG on a rough surface of $TiO_2$ (black line). Reprinted with permission from Ref. [71], [72].

### *3.3.Chemical p-type and n-type doping of graphene*

One of the most prevalent techniques of manipulating the electrical properties of graphene without damage to its structure is chemical doping, which is an essential method for controlling the charge carrier concentration in graphene [64, 73-76]. Doping the graphene sheet results in shifting the Fermi level of graphene and changing its electronic properties. Charge transfer between the graphene and the dopant changes the graphene WF. Depending on the WF of the dopant element, the WF of the doped graphene can be higher (if $WF_{dopant}>WF_{graphene}$) or lower (if $WF_{dopant}<WF_{graphene}$). The WF of graphene could change based on two different mechanisms: shifting the Fermi level (due to graphene band filling) and dipole potential arising from the dopants.

Chemical doping of graphene not only manipulates the WF of graphene, but it could also change the sheet resistance and transparency of the graphene layers [26]. Important research effort on different dopants and their effects on graphene WF has been carried out. Chan et al. theoretically investigated the effects of alkali metals and Au adatoms on the WF of graphene [77]. Based on their results, alkali metal adatoms generate the largest dipole moments and result in the largest decrease in WF of graphene. The use of Au, which is an electron acceptor element, led to an increase in the WF of graphene. Increasing the WF of graphene via the Au ion was also reported [22, 26, 78-80].

Benayad et al. investigated the tuning of the electronic structure and the WF of rGO via Au treatment [78]. Increasing the WF of rGO via Au ions (from 4.5 eV to 4.9 eV) was achieved by extracting electrons from rGO sheets and p-doping of the rGO layers. The Fermi level of rGO sheets was downshifted. Interestingly, the amount of charge transfer from the rGO layers to the Au ion depends on the concentration of Au dopant: increasing the Au concentration results in further increase in the WF of rGO. Due to electron donation to Au ions and enhancement of hole concentration in the rGO layers, the Au-doped rGO layer presented a lower sheet resistance (over 50%) compared to the pristine rGO. Modulating the WF of CVD-grown graphene via Au dopant was also reported by Shi and his colleagues [26]. They presented a simple method for stable Au doping of the graphene layer by immersing the graphene layer in $AuCl_3$ solution. Spontaneous reduction of Au ions led to formation of Au particles on graphene sheets. Increasing the immersion time increased the WF of the graphene layers (up to 0.5 eV) [26, 81]. They suggested the application of Au-doped graphene as a transparent tunable electrode in photovoltaic devices, which can increase the power conversion efficiency.

P-doping of CVD-grown FLG via Au ions and n-doping via viologen dopants was investigated by Shin et al. [22] Based on their investigations, modulating the WF of FLG simultaneously affected other key properties such as sheet resistance, mobility, and sheet carrier density. Au-doping increased the WF of FLG, while viologen-doped FLG presented a lower WF compared to the pristine FLG ($WF_{viologen}$<$WF_{FLG}$). They showed that a change of charge mobility for p-doped FLG was negligible and independent of the type of dopant, while that of n-doped FLG decreased considerably. Moreover, the Raman spectrum of FLG was affected by both the p-doping and n-doping mechanisms. The G and D peaks of the p-doped FLG were upshifted compared to the pristine FLG due to the charge transfer from graphene to Au ions. As a consequence, electron-phonon coupling was improved and led to phonon hardening. In n-doped FLG, the charge transfers from the viologen to FLG resulted in phonon softening and downshifting of the G and D peaks. An enhancement of the WF from 4.5 eV (pristine FLG) to 4.8 eV for Au-doped FLG and a decay to 4.0 eV for viologen-doped FLG were also reported [22].

Decoration of FLG sheets with Au nanoparticles (Au NPs) was reported as an effective method for enhancing the electrical conductivity and tuning the WF of FLG and its application as an electrode for GaN-based LEDs and solar cells (SCs) [79]. Decorated FLG via Au NPs presented a high current injection and electroluminescence for LEDs and enhanced power conversion efficiency for SCs. Introducing the Au NPs in FLG resulted in decrease of both sheet resistance and transparency of the graphene layers. Moreover, modulating the WF of graphene via Au NPs was reported to be dependent on the size of the Au NPs [82]. For instance, researchers showed that reducing the size of the Au NPs from 40 to 5 nm decreased the WF of graphene from 5.76 to 5.35 eV, while its catalyst activity was dramatically increased. Since smaller Au NPs have a larger surface area, the transfer of electrons at the graphene/Au NP interface will be improved due to the quantum confinement effect [82].

Comparative analyses of different p-dopants have been also reported [83]. Different metal chlorides ($AuCl_3$, $IrCl_3$, $MoCl_3$, $OsCl_3$, $PdCl_3$, and $RhCl_3$) were studied to investigate the impact on the WF of FLG. CVD-grown FLG on a Cu substrate was subjected to different dopants, resulting in a decrease in both sheet resistance and transmittance (as was mentioned earlier for p-dopants) [79]. Moreover, after metal chloride doping, the G peak of FLG was shifted to a higher wavenumber, in accordance with charge transfer from graphene to metal ions [22, 83]. UPS data presented an increase in the WF of FLG after metal chloride doping from 4.2 eV (pristine FLG) to 5.0, 4.9, 4.8, 4.68, 5.0, and 5.14 eV for $AuCl_3$, $IrCl_3$, $MoCl_3$, $OsCl_3$, $PdCl_3$, and $RhCl_3$ dopants, respectively.

Although the Au ion was introduced several times as a p-doping agent, thus increasing the WF of graphene, recent studies have revealed that, based on the electronegativity of the anion in different Au complexes, the degree of p-doping would be different [84]. The bond strength between the Au cations and counter anions would determine the degradation of graphene layers. Therefore, good p-dopants for graphene introduce anions with high electronegativity and high bond strength [84].

The effects of alkali metal (n-doping) on decreasing the WF of graphene layers have been studied theoretically [77] and experimentally [74, 85, 86]. Graphene-CNT composites were doped with alkali metal carbonate and heated to produce the alkali metal oxide (Figure 6a) [85]. Decomposition of alkali metal carbonate to low WF alkali metal oxide induced interfacial dipoles that reduced the WF of the graphene-CNT composite. N-doping of alkali metals on graphene layers reduces the transparency of the graphene sheets, while increasing the sheet resistance (Figure 6b) [74]. An increase in concentration of doping solution results in further reduction of the transparency and, subsequently, an enhancement of the sheet resistance. The spontaneous chemical combination between alkali metals and carbon atoms reduces the WF of graphene. Alkali metals with a large atomic number, such as Cs, tend to provide electrons to other materials and exhibit the greatest impact on WF decrease [74, 87, 88]. Moreover, doping the GO layers with alkali metal carbonate (such as $Cs_2CO_3$) showed a decrease in the WF of GO and, subsequently, reversed the hole-extraction ability to electron-extraction, a property to be considered for applications in high-performance SCs and many other optoelectronic devices [86]. Replacing the –COOH functional groups of the GO with –COOCs groups was reported as a reason for reduced WF of GO from 4.5 eV to 4.0 eV.

However, alkali metals were introduced as one of the best dopants for lowering the WF of graphene. Doping graphene with alkali metals, along with other methods (such as electrostatic gating), could decrease the WF to 1 eV [89]. This low WF value was achieved for a large area of SLG deposited via CVD methods. The WF was first decreased by 0.7 eV via an electrostatic gating approach and later reduced to as low as 1 eV by doping the SLG structure with Cs/O in an ultrahigh vacuum (UHV) environment. The electrostatic gating raised the Fermi level, and the Cs/O dopant lowered the vacuum level. As a consequence, ultralow WF materials were obtained for electron emission and energy conversion applications.

The WF of graphene can be also controlled via a doping process using self-assembled monolayers (SAMs), where the WF difference between the graphene and the substrate leads to a variation of the graphene WF. The substrate $SiO_2$, which is a regularly used for applications of graphene as an electrode in optoelectronic devices, intrinsically leads to p-doping of the graphene layer. For optimizing the performance of OFETs, WF of the graphene coating on $SiO_2$ must be controlled. Jo et al. showed that using the SAMs on a $SiO_2$ substrate would tune the WF of graphene from 3.9 to 4.5 eV [15]. $NH_2$-terminated SAMs resulted in n-doped graphene, while the p-doping induced by $SiO_2$ substrates was neutralized via $CH_3$-terminated SAMs [90]. Furthermore, Seo et al. investigated tuning the WF of graphene using heptadecafluoro-1,1,2,2-tetrahydrodecyl-trichlorosilane (HDF-S) as SAMs [91]. HDF-S self-assembled graphene exhibited an increase in WF from 4.56 eV to 5.50 eV, while its electrical and optical properties were stable compared to pristine graphene. The results showed that an increase in the WF of graphene via HDF-S was independent of the substrate; for $Al_2O_3$ and $SiO_2$ substrates, the WF values of the HDF-S self-assembled graphene were similar. Moreover, the WF of the HDF-S self-assembled graphene showed great stability in ambient conditions over a period of 50 days. The transfer of electrons from graphene to HDF-S was reported as a reason for the enhancement of the graphene WF.

In the case of GO, functional groups are one of the important factors that control the WF. Although this topic will be reviewed in depth later, it should be mentioned here that the replacement of functional groups with other chemical compounds would affect the WF. For instance, replacing the –COOH groups of GO with Cs ions was reported as an effective factor for decreasing the WF of GO and reversing its hole-extraction ability to the electron-extraction property [86]. The decoration of GO layers with $–OSO_3H$ groups (sulfated GO or $GO{-}OSO_3H$) enhanced the hole extraction ability of the GO layer and improved its applicability as a hole extraction layer for polymer SCs (PSCs) [92]. Modulating the WF of GO (4.7 eV) to a proper WF for $GO{-}OSO_3H$ (4.8 eV) made it an ideal layer for Ohmic contact with an electron donor polymer. Furthermore, compared to GO, $GO{-}OSO_3H$ exhibited higher conductivity that resulted in higher power conversion efficiency in the resulting PSC devices. In addition to the introduced metal chloride dopants, there are other materials that increase the WF of graphene. Doping the graphene layer with both bis(trifluoromethanesulfonyl)-amide[($(CF_3SO_2)_2NH$)] (TFSA) and chloroform ($CHCl_3$, CF) increased the WF of the graphene layer and reduced the sheet resistance of the layer, while its transparency remained unchanged [93, 94]. Using TFSA as a p-dopant improved the light harvesting ability of doped graphene/n-Si Schottky junction solar devices and increased power conversion efficiencies. Moreover, the hydrophobic nature of the TFSA dopant increased the environmental stability of the fabricated device. Large-scale, low temperature, and stable p-doping of graphene layers with CF successfully increased the WF from 4.61 eV (pristine graphene) to 5.19 eV for CF-doped graphene. Down-shifting the Fermi level relative to the Dirac point was suggested as a reason for the increased WF [94].

In addition to its type and concentration, dopant orientation with respect to the graphene surface also affected the WF (Figure 6c,d) [29]. Christodoulou et al. theoretically showed that depositing the strong molecular acceptor HATCN would tune the WF of graphene from 4.5 eV up to 5.7 eV. Since reorientation causes a larger overall interface dipole compared to the flat-lying acceptor layer, changing the orientation of the HATCN layer as the p-dopant with respect to the graphene surface resulted in increasing the WF of graphene to 5.7 eV.

Another approach for increasing the WF of graphene between 20 and 180 meV (p-doped graphene) was reported by Yen et al. [95]. They showed that introducing boron gaseous vapor along with $CH_4$ during CVD deposition of graphene would result in boron-doped graphene with a tuned WF. The annealing temperature and time were suggested as the two effective factors of controlling the B-doping level of graphene. $FeCl_3$ was also introduced as a dopant for increasing the WF of graphene. Application of graphene as a flexible, conductive, and transparent electrode or circuit in flexible photovoltaic and OLEDs demands a graphene layer with a WF close to that of ITO. In this regard, Bointon et al. manipulated the WF of large area FLG (up to 9 $cm^2$) by intercalating $FeCl_3$ into its layer [96]. The WF of the $FeCl_3$-FLG reported by this group was 5.1 eV (0.3 eV larger than the WF of ITO), indicating $FeCl_3$-FLG as an ideal alternative to ITO for application in optoelectronic devices. Further investigations to increase the WF of graphene were performed by other groups, where the WF was manipulated from 4.5 (pristine graphene) to 5.10 eV by PEDOT chemical doping, as application as the anode in optoelectronic devices [97]. The WF of FLG was also successfully increased by introducing Ag NPs onto the FLG structure, resulting in an increase in WF from 4.39 to 4.55 eV, while its sheet resistance and transmittance were stable [98].

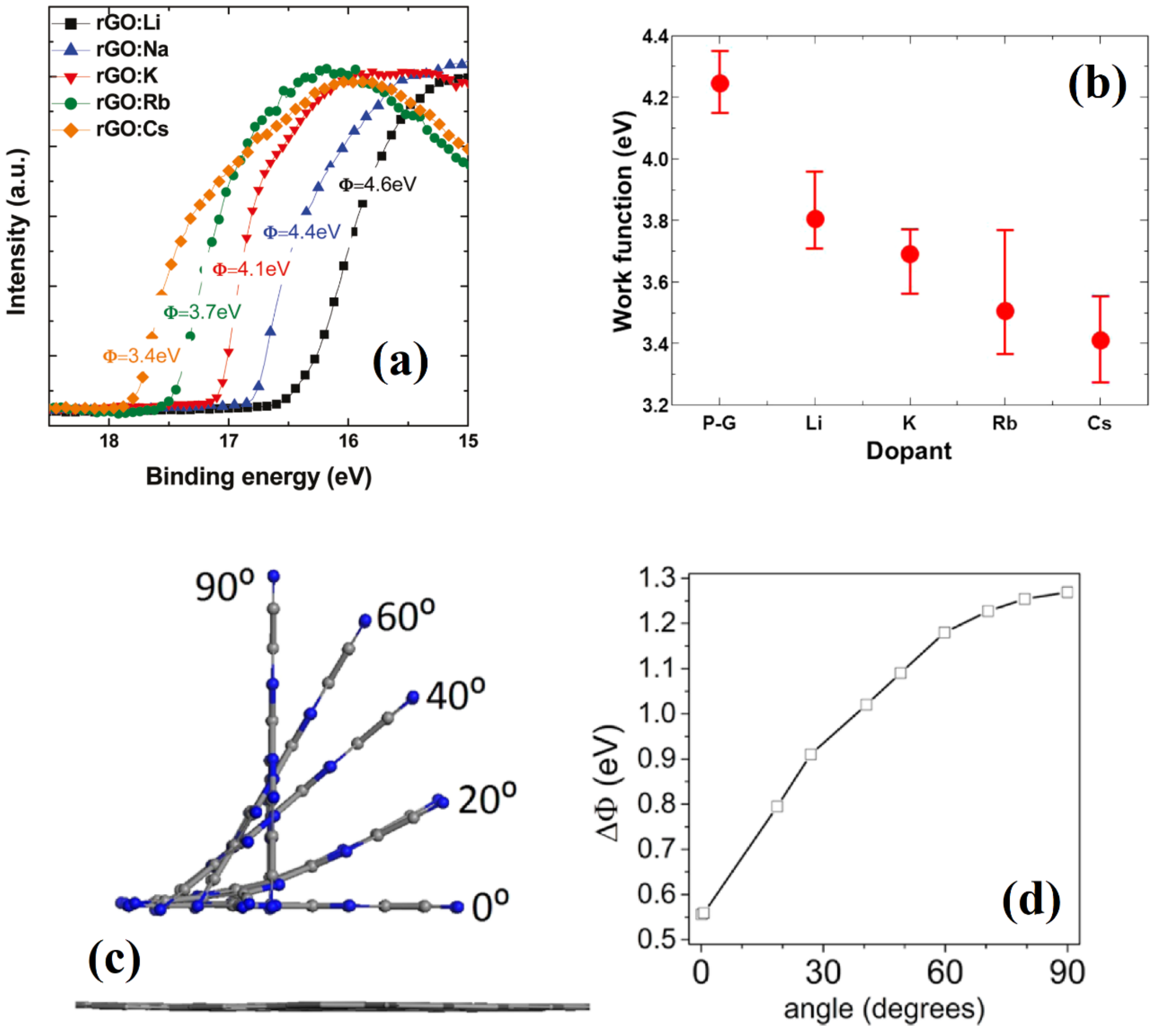

**Figure 6.** Tuning the WF of (a) GO, (b) graphene with different alkali metal dopants, (c) schematic image of HATCN with different configuration respect to graphene, and (d) changing the WF of graphene with HATCN dopant with different configuration. Reprinted with permission from Ref. [99], [100], [101].

For producing low-WF graphene, apart from using alkali metals, other graphene dopants such as PEIE could successfully decrease the WF from 4.6 for pristine graphene to 3.8 eV [102]. Furthermore, functionalization of GO by amino acids was reported as an effective method for reducing the WF of GO from 5.3 eV (pristine GO) to 4.0 eV (amino acid-functionalized GO) [103]. The low cost, abundance in nature, and eco-friendly amino acids were some of the advantages of these functional materials. Additionally, the large molecule dipoles and excellent charge transport properties qualify them as great candidates for modulating the WF of GO. Functionalization of GO (with the tuned WF) has been also reported as an effective dopant for controlling the WF of CVD-grown graphene for its application in optoelectronic industries [104].

Almost all the chemical doping processes of graphene lead to transfer of charge between graphene and dopants or chemisorption of the dopant on graphene layers, which affect the WF and electrical structure of the layers. A theoretical analysis showed that the WF of graphene can be manipulated by physisorption of dopants (without transferring charge between graphene and the dopants) [105]. The results of the DFT calculation suggested that both PEN and PFP organic molecules would shift the vacuum level of graphene without charge transfer in the interface region (PEN as a p-dopant and PFP as an n-dopant). The variation of the WF was explained in terms of a weak dipole formed in the interface of the molecular surface region caused by charge redistribution. Formation of the weak dipole layer shifted the vacuum level -0.06 eV for the PEN dopant and +0.10 eV for PFP dopant.

The number of graphene layers not only affects the WF of graphene (see previous section), but also the chemical doping behavior of the graphene layer. These two processes are coupled. Doping SLG and BLG with electron donating (or withdrawing) chemical compounds in dry conditions showed that SLG is more sensitive to chemical doping than BLG [106]. Chemical doping of both SLG and BLG showed that, after exposure to electron donating (or withdrawing) chemical compounds, the shift of the Fermi level for SLG is larger than that of BLG.

As a summary of this section, Table 1 presents some information on WF engineering of graphene via different chemical dopants. The impacts on electrical and optical properties of the graphene layer induced by the dopants are also shown. Several interesting conclusions can be inferred from the table. First, it is obvious that chemical doping of graphene is a complex task that involves several coupled factors such as type of dopant, concentration of dopant, doping time, size of the dopant (in the case of nanomaterials), orientation of the dopant, and graphene thickness. The range of WF values spans from 3.25 eV for FLG doped with CsF to 5.54 eV for rGO doped with FTS. As shown in Table 1, doping graphene varies the conductivity by several orders of magnitude and the transmittance by about 30%. Such a broad spectrum of properties makes doping an interesting method for controlling the graphene characteristics in a wide range of applications.

Table 1. Effect of chemical doping on electrical and optical properties of graphene

| **Ref.** | **Type of graphene production** | **Type of tuning method** | | **Work function (eV)** | **Method of WF measuring** | **Electrical properties** | **Transmittance (%)** |
|---|---|---|---|---|---|---|---|
| [107] | Graphene (CVD) | Chemical doping | Pristine | 4.5 | UPS | 201 Ω/sq | Not available |
| | | | $AuCl_3$ | 4.8 | | 149 Ω/sq | |
| | | | 2,3-dichloro-5,6-dicyanobenzoquinone (DDQ) | 4.7 | | 190 Ω/sq | |
| | | | 1,1′ dibenzyl-4,4′-bipyridinium dichloride ($BV^{2+}$) | 4.0 | | 260 Ω/sq | |
| [99] | rGO | Chemical n-doping | Pristine | 5.1 | UPS | Not available | Not available |
| | | | $Li_2CO_3$ | 4.6 | | | |
| | | | $Na_2CO_3$ | 4.4 | | | |
| | | | $K_2CO_3$ | 4.1 | | | |
| | | | $Rb_2CO_3$ | 3.7 | | | |
| | | | $Cs_2CO_3$ | 3.4 | | | |
| [100] | FLG (CVD) | Chemical doping | Pristine | 4.25 | UPS | 1100 Ω/sq | 96.7% (at 550 nm) |
| | | | $Li_2CO_3$ | 3.8 | | 2050 Ω/sq | 96.1% (at 550 nm) |
| | | | $K_2CO_3$ | 3.7 | | 1750 Ω/sq | ---- |
| | | | $Rb_2CO_3$ | 3.5 | | 2520 Ω/sq | 94.0% (at 550 nm) |
| | | | $Cs_2CO_3$ | 3.4 | | 1500 Ω/sq | ---- |
| [108] | FLG (CVD) | Chemical p-doping | Pristine | 4.42 | UPS | 1000±98 Ω/sq | 89% (at 400 nm) |
| | | | $AuCl_3$ | 5.12 | | 105±7 Ω/sq | 78% (at 400 nm) |
| [109] | rGO | | Pristine | 4.6 | UPS | ~1200 Ω/sq | Not available |
| | | Chemical n-doping | Hydrazine ($N_2H_4$) | 4.25 | | 300 Ω/sq | 80% (at 550 nm) |
| [110] | Graphene | Chemical n-doping | Pristine | 5.25 | UPS | Not available | Not available |
| | | | $Li_2CO_3$ | 4.47 | | | |
| | | | $Na_2CO_3$ | 4.41 | | | |
| | | | $K_2CO_3$ | 4.33 | | | |
| | | | $Rb_2CO_3$ | 4.16 | | | |
| | | | $Cs_2CO_3$ | 4.03 | | | |
| [111] | SLG (CVD) | Chemical p-doping | Pristine | 4.2 | UPS | 1100 Ω/sq | 96.7% (at 550 nm) |
| | | | $AuCl_3$ | 5.0 | | 500 Ω/sq | 94.5% (at 550 nm) |
| | | | $IrCl_3$ | 4.9 | | 600 Ω/sq | 94.2% (at 550 nm) |
| | | | $MoCl_3$ | 4.8 | | 720 Ω/sq | 86.2% (at 550 nm) |
| | | | $OsCl_3$ | 4.68 | | 700 Ω/sq | 93.1% (at 550 nm) |
| | | | $PdCl_2$ | 5.0 | | 520 Ω/sq | 92.0% (at 550 nm) |
| | | | $RhCl_3$ | 5.14 | | 620 Ω/sq | 89.5% (at 550 nm) |
| [112] | GO | Chemical n-doping | Pristine | 4.7 | KPFM | $3.8\times10^{-3}$ S $m^{-1}$ (conductivity) | Not available |
| | | | $Cs_2CO_3$ | 3.9 | | $5.0\times10^{-3}$ S $m^{-1}$ (conductivity) | |
| [113] | rGO | Chemical p-doping | Pristine | 4.91 | UPS | 55 kΩ/sq | Not available |
| | | | (tridecafluoro-1,1,2,2,-tetrahydrooctyl) trichlorosilane (FTS) ($C_8H_4F_{13}SiCl_3$) | 5.54 | | 35 kΩ/sq | |
| | | | aminopropyl triethoxysilane (APTS) ($C_9H_{23}NO_3Si$) | 4.31 | | 72 kΩ/sq | |
| [114] | SLG (CVD) | Chemical p-doping | Pristine | 4.61 | UPS | 630 Ω/sq | 97.7% (at 550 nm) |
| | | | CF | 5.19 | | 300 Ω/sq | 97.7% (at 550 nm) |
| [115] | Graphene | Chemical p-doping | Pristine | 4.3 | UPS | 800-1100 Ω/sq | 97% (at 550 nm) |
| | | | $Au(OH)_3$ | 4.6 | | 820 Ω/sq | 93% (at 550 nm) |
| | | | $Au_2S$ | 4.8 | | 600 Ω/sq | 86% (at 550 nm) |
| | | | $AuBr_3$ | 5.0 | | 530 Ω/sq | 95% (at 550 nm) |
| | | | $AuCl_3$ | 4.9 | | 300 Ω/sq | 96% (at 550 nm) |
| [116] | SLG (CVD) | Chemical p-doping | Pristine | 4.56-4.15 | KPFM-UPS | 40 Ω/sq | 89.1% (at 700 nm) |
| | | | HDF-S | 5.15 | UPS | 80 Ω/sq | 88.4% (at 700 nm) |
| | | | | 5.51 | KPFM | | |
| [117] | FLG (CVD) | Chemical p-doping | Pristine | 4.6 | KPFM | 20.52 Ω/sq | 77% (in the visible wavelength range) |
| | | | $FeCl_3$ | 5.1 | | | |
| [118] | FLG (CVD) | Chemical n-doping | Pristine | 4.5 | UPS | Not available | Not available |
| | | | CsF | 3.25 | | | |
| | | | $Cs_2CO_3$ | 3.70 | | | |
| | | | PEIE | 4.20 | | | |
| | | Chemical p-doping | PEDOT | 5.10 | | | |

### *3.4. Work function engineering via UV-treatment*

Treatment with UV radiation, also in combination with other methods, can change the WF of graphene and trigger photochemical processes [119-121]. The duration of treatment is a free parameter that can be tuned accordingly to need. UV irradiation of graphene layers has been reported to increase the WF of graphene [119], but this is not the only effect on its properties. For instance, by increasing the oxygen functional groups via UV radiation for a short time, the sheet resistance of the graphene layer increases and its transmittance decreases. These results suggest that the UV-treated graphene can be used as an alternative for PEDOT:PSS in organic photovoltaic cells. Although the UV-treated graphene layer was damaged by UV treatment (shown by Raman spectral data), in ambient conditions, it showed greater stability as a hole extraction layer for organic photovoltaic cells compared to conventional PEDOT:PSS [119].

UV radiation could also be used as a factor for photochemical chlorination of GO. The results showed that photochemical chlorination of GO continuously increased its WF to 5.21 eV (from 4.96 eV). UV irradiation of GO for 2 min from a mercury-vapor lamp in a closed quartz Petri dish with o-dichlorobenzene vapor resulted in chlorinated GO (Cl-GO). By increasing the UV treatment duration, the WF of the Cl-GO remained stable at 5.20 eV after 450 h of storage in a glove box [122].

While UV treatment was introduced as a method for increasing the oxygen functional groups and WF of graphene, the opposite effect has been also reported. A study has shown that UV radiation can produce oxygen desorption from graphene layers and reduce its hole density (changing the carrier concentration) [120]. Since the WF of graphene is dependent on its carrier concentration, UV treatment decreases the WF by changing its carrier concentration (Figure 7a). Moreover, prolonging the treatment time decreases the WF continuously. It can be then concluded that the most important factor in UV treatment of graphene is the duration of irradiation, and increased efforts are necessary to fully understand the process.

### *3.5. Work function engineering via plasma*

Plasma treatment is another factor that can modulate the WF of graphene and control its electron emission ability [123]. Free parameters in plasma treatments, such as type of ions, exposition time, and power, have been used for tuning the WF of graphene. In general, the type of plasma determines the type of dopant (p-doping or n-doping), and the power effects its concentration.

Sherpa et al. reported an increase in the WF of plasma-fluorinated graphene [124, 125]. Creation of an electrical dipole layer with a negatively charged outer surface via plasma treatment was reported, while increasing the WF would oppose electron escape from the surface. UPS characterization suggested that the increase in the WF of graphene was due to adsorption of fluorine atoms. An increase in fluorine concentration resulted in enhancement in the surface dipole moment and, subsequently, an increase in the WF of graphene [125].

Treatment with $O_2$ plasma was also introduced as an effective method for manipulating the WF of graphene [126]. Yang and his coworkers demonstrated an increase in the WF of GO to 5.2 eV. Such a value is much higher than the average WF of GO (~4.96 eV) and could facilitate its application as a hole transport layer in organic SCs. Increasing the WF of GO due to $O_2$ plasma treatment was explained in terms of the higher electronegativity of O atoms. Increasing the electronegativity of the surface of GO increases the surface dipoles, the hole mobility, and the hole extraction [126].

As opposed to $O_2$ plasma, nitrogen plasma treatment decreased the WF of CVD-grown graphene from 4.91 eV to 4.37 eV according to Zeng et al., and the plasma-treated graphene exhibited n-type behavior, while untreated graphene samples presented p-type behavior (Figure 7b) [127]. The WF of graphene was successfully controlled by plasma treatment under different radio-frequency power conditions. Increasing the radio-frequency power led to an increase in both the number of graphitic nitrogen and the electron concentration, thus shifting the Fermi level to higher energies.

The effects of nitrogen plasma treatment on the edges of graphene and at its graphitic site on WF also have been studied [128]. Nitrogen plasma treatment resulted in nitrogen doping of the graphene layer, changing its WF from 4.3 eV to 5.4 eV. The WF of graphene correlated with the duration of plasma treatment (amount of nitrogen doping) and was dependent on the nitrogen-doped sites of graphene. If the graphitic sites of samples were nitrogen-doped, the WF of the sample decreased, while nitrogen doping at pyridinic or pyrrolic sites increased the WF [128].

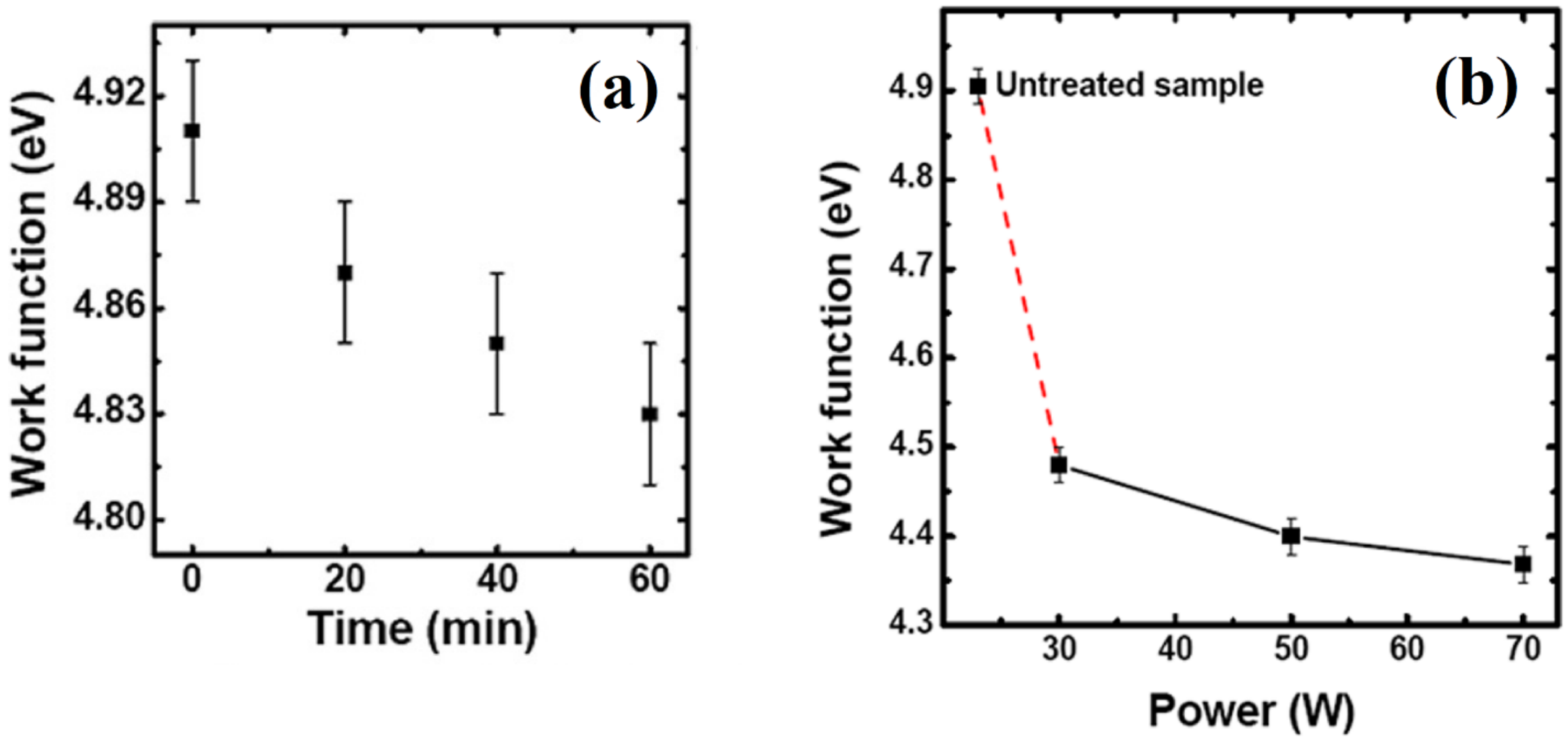

**Figure 7.** Changing the WF of graphene (a) changing the UV radiation time and (b) changing the plasma power. Reprinted with permission from Ref. [129], [130].

### *3.6. The effect of defects and molecular overlayers on the work function of graphene*

One of the biggest obstacles in producing graphene is finding a method for synthesizing defect-free graphene [131, 132]. Defects in the graphene structure limit its application in some specific areas. Moreover, the electronic characteristic of graphene, which is key in most fields of application, is drastically affected by disorder. For this reason, several groups have investigated the effects of disorder, molecular over layers, and grain boundaries on the WF of graphene. There are different types of defects such as carbon vacancies, grain boundaries, wrinkles, Stone-Thrower-Wales defects, and hydrogenated edges that could affect the WF of graphene [27, 133, 134]. Since producing an ideal defect-free graphene layer is impossible, obtaining knowledge on the dependency of the WF on the defect is essential for engineering the defect to control the WF of graphene.

Bae et al. theoretically showed that the WF of a graphene layer with Stone-Thrower-Wales defects was smaller than the WF of pure graphene, while the WF of a graphene layer with a hydrogen-passivated $C_2$ vacancy is higher than the WF of pure graphene (Figure 8a,b) [27]. Long and his co-workers used KPFM to investigate the effects of grain boundaries and wrinkles on the WF of graphene on a $SiO_2$ substrate [133]. KPFM is an interesting diagnostic tool to study defects because, as shown in Sec 2.1, a map of the WF can be obtained. They showed that the doping behavior of the $SiO_2$ substrate was different for different line defects (grain boundaries, standing-collapsed wrinkles, and folded wrinkles) (Figure 8c, d, e). They suggested defect engineering of graphene layers as an effective method for manipulating its WF. The shape and height of the defect were mentioned to explain the difference in WF. In the case of wrinkles, the WF increased along the axial center and decreased at the edges. The results showed a height of 0.5 nm for grain boundaries, 0.84 nm for folded wrinkles, and 1.6-4.6 nm for standing collapsed wrinkles [133].

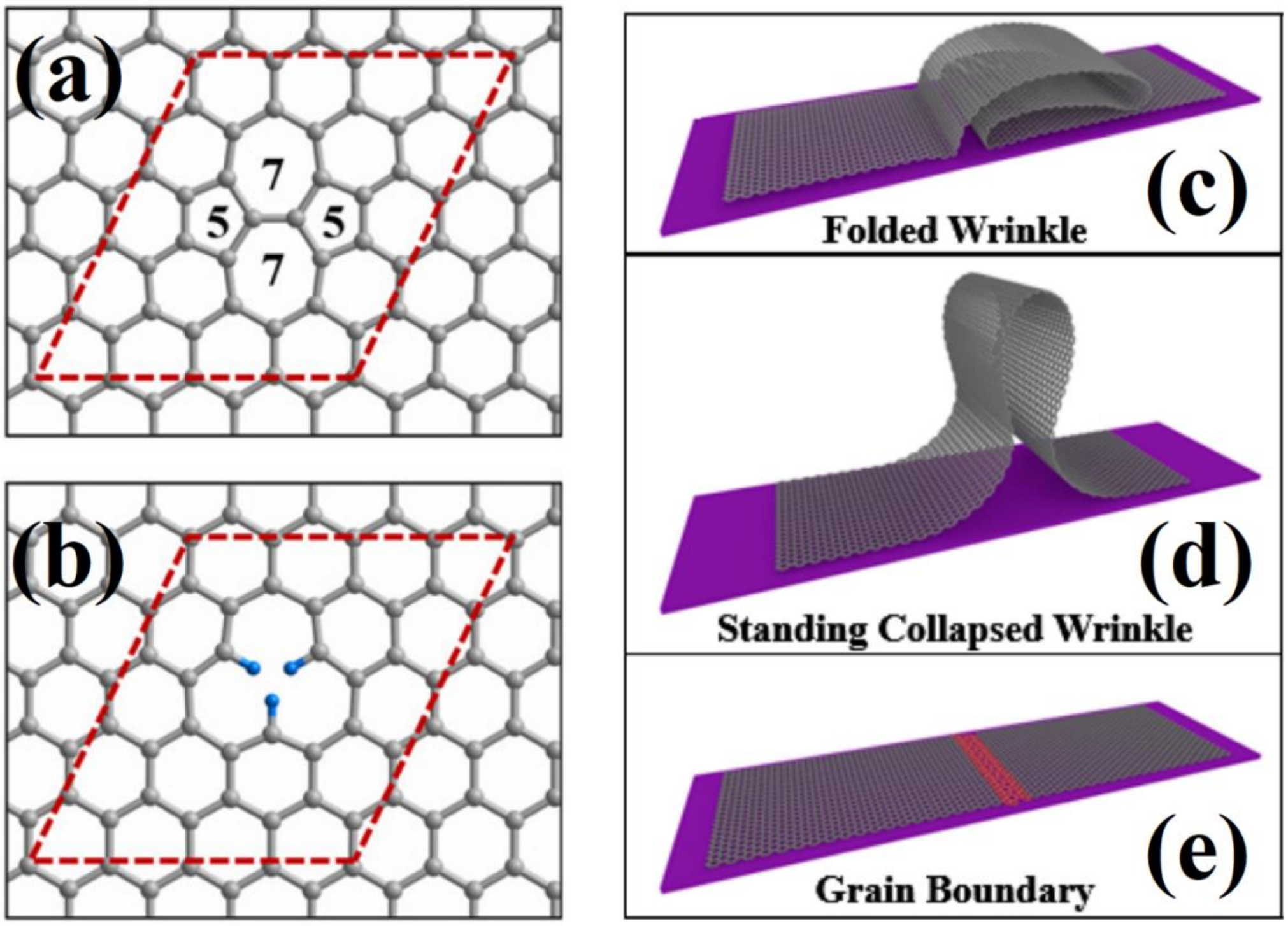

**Figure 8.** Different types of defects and disorders (a) the Stone-Thrower-Wales defect, (b) the SLG with three passivating hydrogen atoms, schematic morphology of (c) folded wrinkle, (d) standing collapsed wrinkle, and (e) grain boundary. Reprinted with permission from Ref. [135], [136].

### *3.7. Effects of functional groups of GO on its work function*

Based on the different oxygen functional groups, GO has emerged as a solution-processable material and a candidate to form a graphene family for applications in research and industry [137-143]. GO could be defined as a graphene layer decorated with carbonyl, hydroxyl, epoxy, and ether groups. Manipulating the concentration of these groups and changing their position on graphene sheets would produce different characteristics of GO [5]. Moreover, by investigating the effects of different functional groups on the WF of GO, the reduction method can be optimized [144]. The combination of different reduction methods [139, 145] and changing the concentration of different functional groups of GO with an appropriate reduction approach (chemical, electrochemical, thermal, etc.) [146] successfully controlled the WF. However, although different functional groups facilitate the application of GO, the impact on electrical, optical, and thermal properties can, in some cases, deteriorate the potential applications of the material. Therefore, reduction of GO (rGO) for changing its functional groups and controlling its electrical and optical properties must be carefully considered [5, 140].

The presence of different oxygen functional groups on the GO surface resulted in difference of its WF compared to graphene [147]. The large electronegativity of O atoms on GO causes surface dipoles via extraction of electrons from graphene, which increases the WF of GO to 4.9 eV. The results of DFT calculation presented by Kumar et al. showed that changing the concentration of carbonyl groups varied the WF of GO from 4.4 to 6.8 eV, which was attributed to the large dipole moment resulting from the C=O double bond (Figure 9a) [146]. Changing the concentration of epoxy groups varied the WF of GO from 4.35 to 5.6 eV, and changing to concentration of hydroxyl groups, showed a change from 4.25 to 4.95 eV. Therefore, the concentration of different functional groups on GO sheets can modulate the WF within a wide range.

Among the different reduction methods of GO, thermal reduction was regularly reported as an effective method in changing the functional groups of GO and recovering the π-conjugated structure of graphene sheets along with enhancing the electrical properties of the GO sheets. In such a reduction method, optimization of the annealing temperature is essential [36, 137, 138]. Liu et al. showed that increasing the annealing temperature removed more functional groups from the GO and increased the conductivity of the rGO sheets while decreasing the WF of rGO from ~5.0 eV (pristine GO) to 4.5 eV (annealed at 230 °C) [138]. The rGO was used as a hole transport layer in bulk heterojunction SCs. The rGO annealed at 230°C exhibited better performance as a hole transport layer for bulk heterojunction SCs compared to rGO annealed at 130°C. This was due to fewer oxygen functional groups on GO thermally reduced at higher temperatures and its better conductivity [138]. The thickness of GO was also reported as an effective factor on the rate of oxygen loss during thermal treatment at a temperature higher than 200°C [139].

In addition to thermal reduction, chemical reduction of GO via hydrazine is one of the prevalent methods for removing the oxygen functional groups of GO and recovering the $\pi$-conjugated structure of graphene sheets. This simple and effective method has been used for lowering the WF of GO from 4.4 eV (GO) to 3.4-3.8 eV (rGO) [148]. The combination of different reduction methods has been also analyzed [139, 145], with the aim of obtaining better control over oxygen group reduction. Sygellou et al. combined hydrazine reduction and subsequent thermal annealing (up to 250°C) in ultrahigh vacuum to obtain a highly reduced GO with the lowest WF of 4.20 eV for rGO [139].

### *3.8. Other factors that affect the work function of graphene*

This review has presented many methods for controlling the WF of graphene. However, it is not exhaustive, and there are other factors that can modify its electrical characteristics and affect the WF directly or indirectly. One of them is the abovementioned electrostatic gating (see Sec. 3.3) that is an active method for adjusting the WF (Figure 9b) [53]. The Fermi level of graphene as 2D material can be controlled by doping, while the Fermi level of regular 3D materials is pinned at the surfaces. For directly lowering its WF, graphene voltage could be biased relative to a gate, resulting in compensating charges. As a result of increasing the population of charge carriers in graphene, the Fermi level shifts from its equilibrium value [53]. Furthermore, electrostatic gating could be combined with other methods to result in ultralow WF materials for electron emission and energy conversion applications [89].

The in-plane orientation of graphene on a substrate can also affect the WF (Figure 9c,d) [46]. Since application of graphene in most cases demands a metal contact, the interaction between graphene and the metal, such as electron transport and the type of graphene/metal contact (Ohmic or Schottky), is significant. Murata et al. showed that growing graphene on Pd (111) would lead to six different in-plane orientations of SLG, where all the WFs were different. Results of LEEM images exhibited that the WF of SLGs with different domain orientations varied by up to 0.15 eV. The study suggested that variation of WF for different graphene domains was increased from the orientation-dependent charge transfer from Pd (111) to SLG. Therefore, control of the electronic characteristic of graphene demands precise control over the relative orientation of the metal and graphene interface [46]. The results of some different WF modulation methods are presented in Table 2.

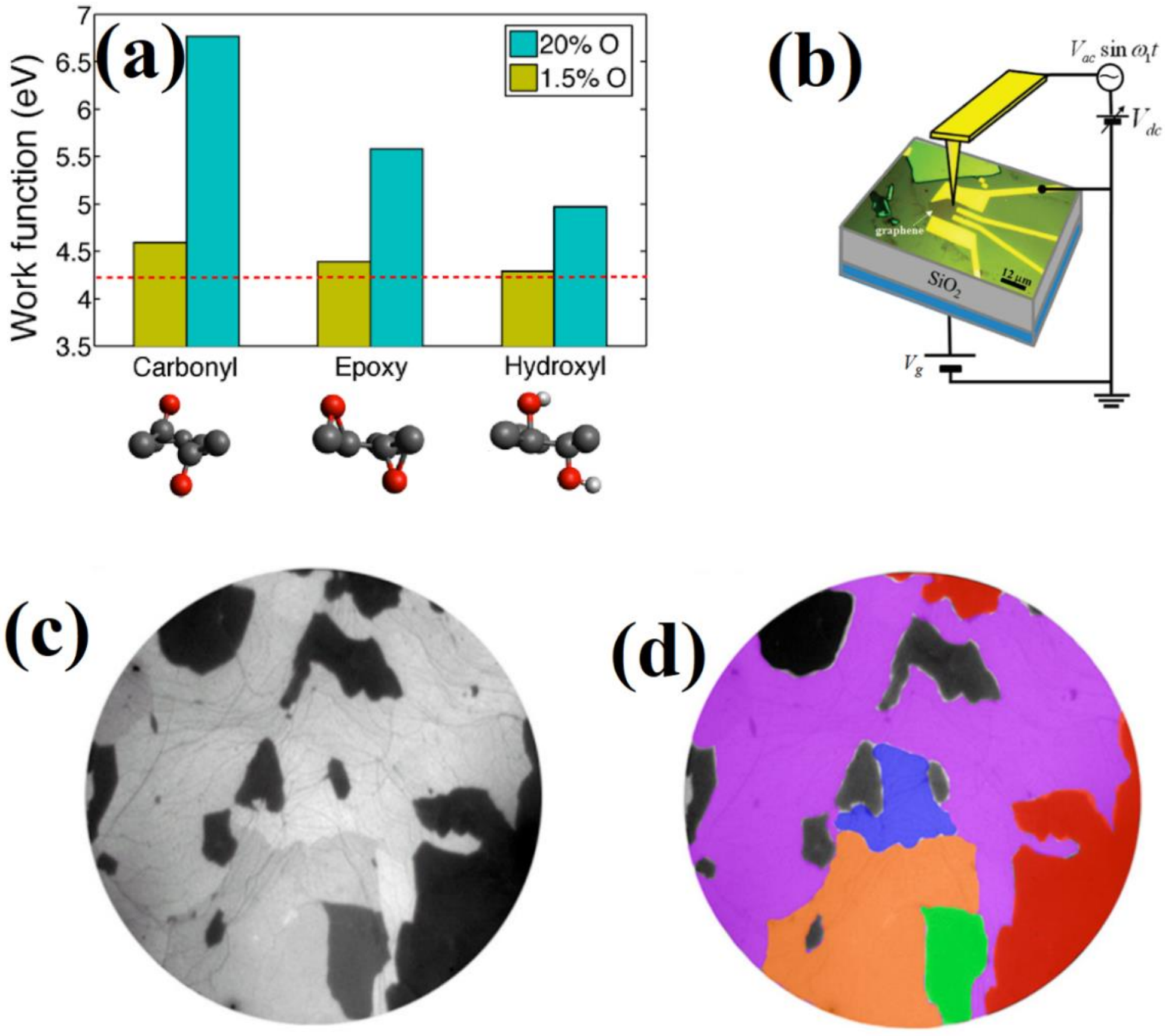

**Figure 9.** (a) Effect of different functional groups on WF of GO for two different oxygen concentration (the dashed red line is the WF of graphene), (b) schematic electrostatic gating of graphene for modulating its WF ($V_g$ is the gate voltage), (c) LEEM image of SLG (with different in-plane orientation) on Pd (111), and (d) six different SLG domains. Reprinted with permission from Ref. [149], [150], [151].

Table 2. Effect of different modulation methods on WF of graphene.

| Ref. | Type of graphene production | The effective factor | | Work function (eV) | Method of WF measuring |
|---|---|---|---|---|---|
| [152] | Epitaxial growth | Thickness | SLG | $WF_{SLG}$ | KPFM |
| | | | BLG | $WF_{SLG}$+(135±9 meV) | |
| [153] | Epitaxial growth | Thickness | SLG | 4.55±0.02 | KPFM |
| | | | BLG | 4.44±0.02 | |
| [67, 154] | Theoretical study | Metal contacts | Pristine | 4.48 | Theoretical study |
| | | | Ni contact | 3.66 | |
| | | | Co contact | 3.78 | |
| | | | Pd contact | 4.03 | |
| | | | Al contact | 4.04 | |
| | | | Ag contact | 4.24 | |
| | | | Cu contact | 4.40 | |
| | | | Au contact | 4.74 | |
| | | | Pt contact | 4.87 | |
| [155] | Mechanical exfoliation | Metal contacts | Pristine | 4.5 | SPCM |
| | | | Ti contact | 4.27 | |
| | | | Au contact | 4.75 | |
| [69] | CVD | Contact | Pristine (MLG) | 4.58±0.08 | KPFM |
| | | | PEO | 4.53±0.03 | |
| | | | $Cs_2CO_3$ | 4.36±0.05 | |
| | | | WPF-6-oxy-F | 4.25±0.03 | |
| [68] | CVD | Metal contact | Pristine (MLG) | 4.40 | UPS |
| | | | Al contact | 3.77-4.40 | |
| [156] | CVD | Metal contact | Pristine | 4.5 | Capacitance-voltage (C-V) |
| | | | Cr/Au contact | 4.3 | |
| | | | Ni contact | 5 | |
| | | | Au contact | 4.62 | |
| | | | Pd contact | 4.6 | |
| [71] | CVD | CNT contact | Pristine | 4.5 | Thermionic emission |
| | | | CNT/graphene | 4.74 | |
| [157] | Modified Hummer method | Reduction method | Pristine GO | 4.7 | UPS |
| | | | Hydrazine hydrate(HYD)-rGO | 4.2 | |
| | | | Thermal-rGO | 4.5 | |
| [158] | CVD | UV/ozone | Pristine | 4.3 | UPS |
| | | | UV/ozone treated | 4.85 | |
| [159] | Highly oriented pyrolytic graphite (HOPG) | $N_2$-plasma | Pristine | 4.3 | UPS |
| | | | Plasma-treated | 5.4 | |
| [150] | Mechanical exfoliation | Electrostatic gating | Pristine | 4.5 | SKPM |
| | | | SLG | 4.5-4.8 | |
| | | | BLG | 4.65-4.75 | |

## 4. Summary and outlook

The extraordinary properties of graphene, including strength, flexibility, transparency, electrical and thermal conductivities, and lightness, make it a good candidate for a wide range of applications that include solar cells, supercapacitors, water desalination, medicine, and many others. Remarkably, for electronic and optoelectronic applications, graphene exhibits the ability to act as both an anode and cathode. In this respect, having control of the WF of graphene is crucial because it critically affects the performance of the devices. The results summarized in this review reveal an important effort toward not only tuning the WF, but also upon understanding the dominant physical mechanisms and the collateral impact on other properties of graphene.

The abundance of available methods presented in this work, such as the number of layers, metal and insulator contacts, doping, defects, functional groups, and UV and plasma treatments, show that it is possible to control the WF of graphene from about 3.25 eV to 5.54 eV. Such an interval, with a width of 2.3 eV, allows broad margins for applications because the performance is typically very sensitive to the WF. The progress on tuning methods took advantage of the development of WF characterization techniques that, as shown in this review, are currently dominated by UPS and KPFM. In addition to the WF of bulk samples, it is also possible to obtain maps of the WF distribution throughout the material, revealing noteworthy information about the homogeneity. Interestingly, most of the tuning methods can be controlled during the synthesis of the graphene (number of layers, defects, functional groups). As a consequence, a precise control on synthesis methods emerged as a key factor on varying its WF. For instance, taking advantage of beneficial methods with the possibility of controlling the number of graphene layers in a vacuum environment (without contamination), such as CVD, would be effective to engineer the produced graphene [132, 160].

Since industrial applications are commonly driven by trade-off analyses involving cost and performance, special attention has been dedicated in this review to GO and rGO. It is noteworthy that specific techniques have been developed. The results demonstrated that increasing the number of graphene layers increases the WF and affects the reduction of GO structure. The type of functional groups and their concentration on the GO layers, as well as doping with alkali metals, are effective procedures for controlling the WF. Studies have also shown that synthesis technique and the reduction method must be carefully monitored. Using a combination of different reduction methods for removing the functional groups would be very effective for controlling the amount of oxygen functional groups of the final products.

To date, there are important obstacles related to the production and application of conventional metal oxides. The use of graphene could certainly substitute for these materials, but the synthesis of high-quality graphene with a large surface area presents production issues that must be overcome.  If these problems could be solved, graphene could be applied (as both anode and cathode) in future flexible optoelectronic devices.

**Acknowledgment**

This work was supported by Agencia Estatal de Investigación (Ministerio de Ciencia, Innovación y Universidades of Spain) under the project ESP2017-82092-ERC (AEI). GSA work is supported by the Ministerio de Ciencia, Innovación y Universidades of Spain under the Grant RYC-2014-15357.